\documentclass[twocolumn,aps,prd,floats,floatfix,showpacs,superscriptaddress,nofootinbib]{revtex4}
\usepackage{amsmath,amsfonts}
\usepackage{graphicx}
\begin{document}
\title{ General Relativistic  versus Newtonian: a universality in radiation hydrodynamics}
\author{Edward Malec}
\affiliation{Instytut Fizyki Mariana  Smoluchowskiego,  Instytut Fizyki, Uniwersytet Jagiello\'nski,Reymonta 4, 30-059 Krak\'{o}w, Poland }
\affiliation{Physics Department, University College, Cork, Ireland}
\author{Tomasz Rembiasz $^{1,}$}
%\affiliation{Instytut Fizyki Mariana  Smoluchowskiego,  Instytut Fizyki, Uniwersytet Jagiello\'nski,Reymonta 4, 30-059 %Krak\'{o}w, Poland}
\affiliation{Max-Planck-Institut f\"{u}r  Astrophysik, Garching, Germany}

\begin{abstract}
We compare Newtonian and general relativistic descriptions of the stationary accretion of self-gravitating fluids onto compact bodies.  Spherical symmetry and thin gas approximation are assumed.  Luminosity depends, amongst other factors,    on the  temperature and the contribution of gas to the total mass, in both -- general relativistic ($L_{GR}$) and Newtonian ($L_N$) -- models. We discover a remarkable universal behaviour for   transonic flows: the ratio of respective luminosities  $L_{GR}/L_N$ is independent of the fractional mass of the gas and  depends  on asymptotic temperature.  It is close to 1 in the regime of low asymptotic temperatures and can grow by one order of magnitude  for high temperatures. These conclusions are valid for a wide range of polytropic equations of state.
\end{abstract}

\maketitle

\section{Introduction}
 
The Newtonian description of accretion has many advantages over the general relativistic modelling; it is much simpler conceptually,  analytically and numerically.  Therefore it is of interest to specify limits of the validity of Newtonian models. It is already known, that they can fail  in  the   accretion of transonic  test  fluids \cite{EM99}-\cite{Kinasiewicz}.   It is not suprising  that the Newtonian description is not   suitable for a class of  transonic flows in radiation hydrodynamics.    But it is suprising that there emerges a universality unknown in the existing literature.  We prove that the ratio  $L_{GR}/L_N$ of general relativistic $L_{GR}$ and Newtonian  $L_N$ luminosities   is sensitive only to the asymptotic temperature, assuming the thin gas approximation and some additional   conditions.
  
Models considered here constitute extensions -- Newtonian and general relativistic -- of  the classical Bondi model \cite{Bondi}  to radiation hydrodynamics.  They have been formulated by several authors (incomplete list includes \cite{Shakura} -- \cite{Karkowski2009}; see also references therein). We make a number of  assumptions  --- spherical symmetry, a polytropic equation of state and the thin gas  approximation in the transport equation   \cite{Mihalas}. It is assumed that the accretion is quasi-stationary. 

There are two kinds of possible general relativistic effects.   The first is related to the  backreaction -- and that includes  self-gravity and dependence on the fractional mass  content (defined later as $1-y$) of the accreting gas -- and the other  is related to the asymptotic speed of sound $a_\infty $ (we shall occasionally use   the term  "asymptotic temperature"; both quantities are proportional).   The same boundary conditions are assumed  in both models. Each of the two quantities,   $L_{GR}$ and $L_N$, separately depends on $y$ and  $a_\infty $, but their ratio $L_{GR}/L_N$ depends only on the asymptotic temperature.   This   universality   means that the  enhancement  of $L_{GR}/L_N$ can be found by solving  the accretion of test fluids in hydrodynamics without radiation. This is a much simpler -- algebraic -- problem than the original one. These facts are valid for  polytropic equations of state $p=K\rho_0^\Gamma $ with $1<\Gamma <5/3$. 

The order of the rest of this paper is as follows. Section II discusses notation and equations.  Section III shows how the validity of the thin gas approximation constrains  the choice of boundary data. We describe  the form of boundary data for the accretion problem.  Section IV discusses the notion of sonic points and transonic flows in radiation hydrodynamics.  In the first part of Section V we formulate a low-radiation condition  and prove the  universality of $L_{GR}/L_N$. The luminosity of hot accreting gases in a general relativistic model can be much higher than that given  by the related Newtonian counterpart. In the second part of Section V we find a sector of luminous transonic flows which again gives a universal ratio, but now the asymptotic temperature is low and $L_{GR}/L_N\approx 1$.   Section VI describes the numerical scheme that is used in this work, gives the equation of state and  details numerical values for most of boundary data. Section VII reviews numerical results.  The last Section summarizes the main
 conclusions.

\section{Formalism and equations}

\subsection{General relativistic accretion.}  

The metric
\begin{equation}
ds^2=-N^2dt^2+\hat adr^2 +R^2\left( d\theta^2 +\sin^2 (\theta )d\phi^2\right)
\label{1}
\end{equation}
uses comoving coordinates $t, r, 0\le \theta \le \pi , 0\le \phi < 2\pi$: time, coordinate radius and two angle variables, respectively. $R$ denotes the areal radius and $N$ is the lapse.  The radial velocity of gas is given by $U = \frac{1}{N} \frac{dR}{dt}$.  We assume relativistic units with $G=c=1$.
 
The energy-momentum tensor reads $T_{\mu \nu }=T_{\mu \nu }^B +T_{\mu \nu }^E$, where the baryonic part is given by $T_{\mu \nu }^B =\left( \rho +p \right) U_\mu U_\nu +pg_{\mu \nu }$ with the time-like   normalized four-velocity $U_\mu $, $U_\mu U^\mu =-1$. The radiation part   $T_{\mu \nu }^E$  possesses   only four non-zero components, 
$T_0^{0E}\equiv -\rho^E=-T_r^{rE}$ and $T^E_{r0}= T^E_{0r}$. A comoving observer would measure local mass densities, the material density $\rho =T^{B\mu \nu }U_\mu U_\nu $  and the radiation density $\rho^E$, respectively. The baryonic current reads   $j^\mu \equiv \rho_0 U^\mu$, where $\rho_0$ is the baryonic mass density. Its conservation is expressed by the equation
\begin{equation}
\nabla_\mu j^\mu = 0.
\label{2}
\end{equation}
Let   $n_\mu$ be the unit normal to a  coordinate sphere lying in the hypersurface $t=const$ and let $k$ be the related mean curvature scalar, $k={R\over 2}\nabla_i n^i=\frac{1}{\sqrt{\hat a}}\partial_rR$.  The quantity  $j = U_\mu n^\nu NT^{\mu E}_\nu /\sqrt{\hat a} = NT^{0E}_r /\sqrt{\hat a}$ is interpreted as the comoving radiation 
flux density. We assume  the polytropic equation of state $p=K\rho_0^\Gamma $, with constants $K$ and $\Gamma $. The  internal energy density $h$ and the rest and baryonic  mass densities are related by $\rho =\rho_0+h$, where   $h=p/(\Gamma -1)$.
  
There are four conservation equations that originate from  the contracted Bianchi identities, $\nabla_\mu T^{\mu B}_\nu =-\nabla_\mu T^{\mu E}_\nu =F_\nu$ (here $\nu = 0, r$). The radiation force density $F_\nu $   describes the interaction between baryons and radiation. This   formulation of general relativistic radiation hydrodynamics agrees with that of Park \cite{Park}, Miller and Rezzola \cite{Rezzola} and (on the fixed, Schwarzschildean, background) Thorne \textsl{et.\ al} \cite{ThorneF}.

One can find the mean curvature $k$  from the Einstein  constraint equations $G_{\mu 0}=8\pi T_{\mu 0}$  (\cite{EM99}, \cite{Iriondo})
\begin{eqnarray}
&& k = \sqrt{1-\frac{2m(R)}{R} +U^2},
\label{3}
\end{eqnarray}
where   $m(R)$ is the quasilocal mass,  
\begin{equation}
m(R)=M-4\pi \int_R^{R_\infty }dr r^2\left( \rho +\rho^E + \frac{Uj}{k}\right) .
\label{4}
\end{equation}
The integration in (\ref{4}) extends from $R$ to the outer boundary $R_\infty $  of the ball of gas. Its external boundary is connected  to the Schwarzschild vacuum spacetime by a transient zone of a negligible    mass. Thus the asymptotic mass $M$ is approximately equal to $m\left( R_\infty \right)$.  

In the polar gauge foliation one has a new time $t_S(t,r)$ with $\partial_{t_S}=\partial_t-NU\partial_R$. The quantity $4\pi Nk R^2\left( j\left( 1+ \left( \frac{U}{k}\right)^2\right) +2U\rho^E/k\right)$ is the radiation flux measured by an observer located at $R$ in coordinates $(t_S,R)$.  One can show that 
\begin{eqnarray}
\partial_{t_S} m(R)&=&
-4 \pi  Nk R^2\left( j\left( 1+ \left( \frac{U}{k}\right)^2\right)
+ 2\rho^E{U\over k} \right) -\nonumber\\
&&4\pi NU R^2 \left( \rho +p \right)  .
\label{5}
\end{eqnarray}
The mass contained in the annulus $(R, R_\infty )$ changes if the fluxes on the right hand side, one directed outward and the other inward, do not cancel.  The local baryonic flux reads $\dot M=-4\pi   UR^2\rho_0$, it is not constant ($\partial_R\dot M \ne 0$)  and its boundary value reads $\dot M_\infty $.

The accretion process is said to be  stationary (or quasi-stationary) if all relevant physically observables, that are  measured at a fixed areal radius $R$, remain approximately constant during time intervals much smaller than the  runaway instability time scale  $T=M/\dot M_\infty $.  That means that  $\partial_{t_S}X\equiv (\partial_t -NU\partial_R) X = 0$ for $X=\rho_0, \rho, j, U\ldots   $.
 
The above assumptions imply that  in the thin gas approximation    $F_0=0$ and the  radiation force density has only one non-zero component  $F_r=\kappa k\sqrt{\hat a}\rho_0 j$ \cite{Karkowski2009}. Baryons and radiation interact through the elastic Thomson scattering. $\kappa $ is a material constant, in standard units $\kappa =\sigma /\left(  m_p c \right)$ and $c$,  $\sigma $ and $m_p$ are respectively  the speed of light, the Thomson cross section and   the proton mass.

The full system of equations in a form suitable for numerics has been obtained in \cite{Karkowski2009}. It consists of:
\begin{enumerate}
\renewcommand{\theenumi}{\roman{enumi}}
\renewcommand{\labelenumi}{\theenumi )}
\item the total energy conservation
\begin{equation}
\dot M N \frac{\Gamma-1}{\Gamma-1-a^2}+ 2 \dot M N \frac{\rho^E}{\rho_0} =
4\pi R^2 j N k \left( 1+ \frac{U^2}{k^2}\right) +C ;
\label{6}
\end{equation}
the constant $C$ is the asymptotic energy flux inflowing through the sphere of a radius $R_\infty $.

\item The local  radiation energy conservation
\begin{eqnarray}
 &&\left( 1- \frac{2m(R)}{R} \right) \frac{N}{R^2} \frac{d}{dR}
\left( R^2\rho^E\right) =\nonumber\\
&&-\kappa k^2N j\rho_0 + 
2N\left( U \rho^E -kj\right) \frac{dU}{dR} + \nonumber\\
&& 2k\left( jU -k\rho^E\right) \frac{dN}{dR} + 8\pi NR
\left( j^2 - j \rho^E \frac{U}{k}\right) .
\label{7}
\end{eqnarray}

\item  The relativistic Euler equation  (below $a=\sqrt{dp\over d\rho }$ is the speed of sound)
\begin{eqnarray}
&&\frac{d}{dR} \ln a^2 = -\frac{\Gamma-1-a^2}{a^2- \frac{U^2}{k^2}}
\times \nonumber\\
&&\Biggl[ \frac{1}{k^2 R} \left( \frac{m(R)}{R} -2U^2 + 4\pi R^2
\left( \rho_E + p + j \frac{U}{k} \right) \right) - \nonumber\\
&&\kappa j \left( 1- \frac{a^2}{\Gamma-1} \right) \Biggr] .
\label{8}
\end{eqnarray}
\item The baryonic mass conservation
\begin{equation}
\frac{dU}{dR} = -\frac{U}{\Gamma -1 -a^2} \frac{d}{dR} \ln a^2 - \frac{2U}{R} +
\frac{4\pi R j}{k}.
\label{9}
\end{equation}
\item The equation for the lapse
\begin{equation}
\frac{dN}{dR}=N \left( \kappa j \frac{\Gamma-1-a^2}{\Gamma-1} + \frac{d}{dR}
\ln \left( \Gamma -1 -a^2 \right) \right) .
\label{10}
\end{equation}
\end{enumerate}

Equations (\ref{3}), (\ref{4}) and (\ref{6} -- \ref{10}) give the complete model used in numerical calculations.  

The asymptotic data for the accretion must satisfy several physical conditions. We assume the inequalities  $a^2_\infty \gg M/R_\infty \gg U^2_\infty$ ensuring, as demonstrated by Karkowski et al. \cite{AA} and Mach et al. \cite{Mach}, that the assumption of stationary accretion is reasonably well satisfied. They   are probably required by the demand of stability (see a discussion in \cite{AA} and studies of stability of accreting flows in Newtonian hydrodynamics \cite{Mach}). In the asymptotic region  $j_\infty \approx \rho^E_\infty$ and the total luminosity is well approximated   by $L_0=4\pi R^2_\infty j_\infty $. The total luminosity is related to the asymptotic accretion rate $\dot M_\infty $ by  \cite{Karkowski2009}
\begin{equation}
L_0= \alpha \dot M_\infty \equiv \left(  1- \frac{N\left( R_0\right)}{k\left( R_0\right)}
\sqrt{1 - \frac{2m\left( R_0\right)}{R_0}} \right) \dot M_\infty .
\label{11}
\end{equation}
Here $R_0$ is the size of the compact core and the quantity  $\alpha$ can be interpreted as a   binding energy per unit mass.  

\subsection{Newtonian  approximation}

 The notation is as in the preceding part of this  section. The mass accretion flux is now R-independent, in contrast to the general relativistic case, $\partial_R \dot M =0$ and the baryonic mass density $\rho_0$ coincides with $\rho $.  $\phi (R)$ is the Newtonian gravitational potential,
\begin{equation}
\phi (R) = -{M(R)\over R} -4\pi \int_R^{R_\infty } r \rho (r) dr;
\label{12}
\end{equation}
$M(R)\equiv M- 4\pi \int_R^{R_\infty } r^2 \rho (r)  dr$ is the mass contained within the sphere $R$.
 
The Newtonian model can be described by two basic equations \cite{AA}:

i) the energy conservation equation
\begin{eqnarray}
&&L_0 - L(R) = \nonumber\\
&& \dot M \left( \frac{a^2_\infty}{\Gamma - 1} +\frac{U^2_\infty}{2} + \phi (\infty ) -
\frac{a^2}{\Gamma - 1} - \frac{U^2}{2} - \phi (R) \right) .\nonumber\\
&&
\label{13}
\end{eqnarray}
ii) The luminosity equation
\begin{equation}
L = L_0\exp \left( \frac{-\kappa  \dot M }{  4\pi R}\right) =L_0\exp \left( \frac{-L_0 \tilde R_0}{L_E R}\right) .
\label{14}
\end{equation}
Notice that the luminosity has the same form as in the case of test fluids \cite{Shakura}. Here we introduced the Eddington luminosity $L_E = 4\pi M/\kappa$ while $\tilde R_0 \equiv GM/|\phi (R_0)|$ is a kind of modified size measure of the compact body. In the case of test fluids $\tilde R_0=R_0$.   We assume $L_0=|\phi \left( R_0\right) |\dot M$; notice, however,  that for small $\alpha $ this relation (with $\alpha = |\phi (R_0)|$) appears as the Newtonian limit of Eq. (\ref{11}).

\section{Thin gas approximation and boundary data}

The thin gas approximation demands   that the optical thicknes \cite{Mihalas} of the cloud is smaller than one, i.e.
\begin{equation}  
\tau = \int^{R_\infty}_{R_0} n(r) \sigma dr  < 1,
\label{thin}
\end{equation}
where $n$ is the baryonic number density. Notice that $n={\rho_0\over m_p}$ if we assume the monoatomic hydrogenic gas. Assuming that $n$ decreases  to the asymptotic value $n_\infty $, we arrive at $1 > R_\infty  n_\infty \sigma $. Thus the rough condition for the validity of the thin gas approximation  is that the radiation free path $l\equiv 1/(n_\infty \sigma )$ is not shorter than  the size of the cloud, $l  >R_\infty $.     This implies       $\rho_0 <{m_p\over R_\infty \sigma }$ and (taking into account that $\rho_0\approx \rho $) estimates the mass of  gas,
 $M_g < {4\pi \over 3\kappa c}R^2_\infty $. Denote the solar mass by $ M_{\odot}$  and define $10^s\equiv {R_\infty \over M}$.  One obtains an estimate consistent with the thin gas approximation
\begin{equation}
{M_g\over M} <  10^{-21} \times 10^{2s}\times {M\over  M_{\odot}}.
\label{15}
\end{equation}
We  choose    $s$ and $M$  that give the right hand side of (\ref{15}) of the order of unity. In such a case  a significant part of the total mass $M$ would be contributed by the gas itself. That could allow for the strong impact of backreaction and selfgravitation onto accretion. It is clear that there is a scaling freedom  -- one can trade the size (represented by the exponent $s$) for the total mass without changing the bound in (\ref{15}).

The boundary data set is the same for the Newtonian and general relativistic models. Thus we specify in both cases the same values of asymptotic masses $M$, masses of the core, the binding energy per unit mass $\alpha =|\phi \left( R_0\right) | $, the asymptotic speed of sound $a_\infty $ and  the  size $R_\infty $.   The total luminosity $L_0$ is not a free data, but it  results from equations. We assume identical equations of state  in the two models.
%%%%%%%%%%%%%%%%%%%%%%%%%%%%%%%%%%%%%%%%%%%%%%%%%%%%%%%%%%%%%%%%%%%%%%%%%%%
\section{Sonic points and luminosity}

We shall study  transonic flows. For these flows there exists a radius $R_\ast$ such that $a_\ast=|{\vec U_\ast }|$; the speed of sound is equal to the length of the spatial part of the velocity vector.   Henceforth all quantities denoted by asterisk  will   refer   to a sonic point.

It is clear from the inspection of   equations  that the regularity of solutions demands a particular relation  for the fraction $m_\ast /R_\ast $; here $m_\ast $ is the mass within the sonic sphere.
In the Newtonian model  the three characteristics, $a_{\ast N}$, $U_{\ast N}$, and $m_{\ast N} / R_{\ast N} $ are related as below \cite{AA}
\begin{eqnarray}
a_{\ast N}^2 &=& U_{\ast N}^2 = \frac{ m_{\ast N} }{2 R_{\ast N} }
	\left(1- \frac{L_{\ast N} \kappa }{4\pi  m_{\ast N}  } \right) = \nonumber\\
&&
\frac{  m_{\ast  N} }{2R_{\ast N} }\left(1- \frac{L_{\ast N} M}{L_E m_{\ast N}} \right) .
\label{16}
\end{eqnarray}
In the last equation appears the Eddington luminosity $L_E $. It is clear   that the necessary condition for the  critical Newtonian flow  -- that is, possessing a sonic point -- reads
\begin{equation}
\frac{L_{\ast N}  M}{L_E m_{\ast N} } < 1 .
\label{17}
\end{equation}
Define $x\equiv L_0/L_E$ and $y\equiv m_\ast /M$. Since $L_\ast \le L_0$, the inequality $x<y$ becomes the  necessary condition for a sonic point.  In the general relativistic model,  at the sonic point  $a^2=\frac{U^2}{k^2}$; the denominator of the right hand side  of Eq. (\ref{8}) vanishes and that  implies  the vanishing of the numerator. One obtains
\begin{eqnarray}
&&\frac{1}{k^2 R} \left( \frac{m}{R} -2 U^2 + 4\pi R^2 \left( \rho_E + p +
j \frac{U}{k} \right) \right) =\nonumber\\
&& \kappa j \left( 1-
\frac{a^2}{\Gamma-1}\right).
\label{17a}
\end{eqnarray}
Let us remark, that in the Newtonian limit $a^2\ll 1$  and $4\pi R^2_{\ast GR} \left( \rho_{\ast E} + p_\ast  +
j \frac{U_\ast }{k_\ast } \right) \ll {m_{\ast GR}\over R_{\ast GR}}$. Therefore, in this limit   Eq. (\ref{17a})  coincides with Eq. (\ref{16}). It is obvious that radiation pushes the sonic point inward; if the size of a compact object is bigger than the value  of $R_\ast $ predicted by (\ref{17}) and (\ref{17a}), then the flow becomes subsonic. 
%%%%%%%%%%%%%%%%%%%%%%%%%%%%%%%%%%%%%%%%%%%%%%%%%%%%%%%%%%%%%%%%%%%%%%%%%%
\section{Universality in $L_{GR}/L_N$}

We assume in this Section the polytropic equation of state $p=Kn^\Gamma $ with   $\Gamma <5/3-\epsilon $,  for some small $\epsilon >0$. This restriction is due to the peculiar character of the equation of state corresponding to $\Gamma =5/3$ The constancy of $L_{GR}/L_N$ is valid for all   $\Gamma $'s, although the specific value of this ratio  depends on the equation of state.

\subsection{Low luminosities}

The natural reference quantity for radiating systems is the Eddington luminosity $L_E$. It can be roughly described as the luminosity at which the  infall of gas is prevented.  Thus one might define  weakly radiating systems as radiating with a luminosity $L_0$ (herein $L_0=L_{GR}$ or $L_0=L_N$) that is much smaller than the Eddington luminosity,  $L_0\ll L_E$. We will adopt a different definition, for reasons that will become clear.

{\bf The (XY) condition.}    We will say that an accretion system satisfies the (XY) condition if  $x\ \ll y $.

Notice the trivial fact that $y< 1$. If (XY) holds, that is $x\ll y$, then obviously $L_0\ll L_E$.
Thus the (XY) assumption is stronger than just the statement $L_0\ll L_E$. Another interesting fact is  that (XY) guarantees that the characteristics of the sonic point are essentially unchanged by the radiation  -- see Eqs. (\ref{16}) and (\ref{17a}).   The luminosity  is     the product of $\alpha $ by the (asymptotic) mass accretion rate, and since the mass accretion rate can be formulated completely in terms of the sonic point parameters, it becomes luminosity independent if $x\ll y$.   The general relativistic  mass accretion rate  $\dot M_{GR}$ within the steadily accreting fluid can be expressed as below (see Eq. (6.1) in \cite{EM99})
\begin{eqnarray}
\dot M_{GR} &=& \pi m_{\ast GR}^2 \rho_{\infty GR} \frac{R^2_{\ast GR}}{m_{\ast GR}^2} \left( \frac{a_{\ast GR}^2}{a_\infty^2}
\right)^\frac{(5 - 3 \Gamma)}{2(\Gamma - 1)} \left(1 + \frac{a^2_{\ast GR}}{\Gamma } \right) \nonumber\\
&&\times \frac{1 + 3 a_{\ast GR}^2}{a^3_\infty}.
\label{18}
\end{eqnarray}
The corresponding Newtonian expression reads
\begin{equation}
\dot M_N = \pi m_{\ast N}^2 \rho_{\infty N} \frac{R^2_{\ast N}}{m_{\ast N}^2}
\left( \frac{a_{\ast N}^2}{a_\infty^2}\right)^\frac{(5 - 3 \Gamma )}{2(\Gamma - 1)}    \frac{1  }{a^3_\infty }.
\label{19}
\end{equation}
Eq. (\ref{19}) has been derived  by Kinasiewicz in \cite{KinasiewiczPhd}, but it follows also  from (\ref{18})
in the limit of small sound speeds, $a_{\ast GR} <<1$. The way of writing  these two expressions is not accidental. It has been shown in \cite{PRD2006} that characteristics of the sonic point -- $a^2_{\ast GR}$ and $  \frac{R^2_{\ast GR}}{m_{\ast GR}^2}$ -- do not depend on the fraction of mass carried by the gas. These quantities are dictated just by the asymptotic speed of sound $a_\infty $ in a test fluid model. An analogous result holds in the Newtonian model, as shown in \cite{KinasiewiczPhd}. Therefore  $L_{GR}/L_N =\dot M_{GR}/\dot M_N$    is equal to the ratio $F\times \frac{m_{\ast GR}^2 \rho_{\infty GR}}{m_{\ast N}^2 \rho_{\infty N}}$, where the coefficient $F$ depends on $\Gamma $ (and thus on the equation of state) and on the sonic point parameters $a_{\ast N}$, $a_{\ast GR}$,   $\frac{R^2_{\ast GR}}{m_{\ast GR}^2}$ and $\frac{R^2_{\ast N}}{m_{\ast N}^2}$. Therefore the coefficient $F$  is independent of the mass fraction $y$. Now the masses are approximately equal, $m_{\ast GR}\approx m_{\ast N}$; this is because the masses within the sonic point are well approximated by the masses of the cores, and the latter are equal by definition. The equality of  masses of the cores in both  models is one of our boundary conditions. The asymptotic gas densities $\rho_{\infty GR}$ and $\rho_{\infty N}$ are approximately equal to  $\left( M-m_{\ast GR}\right) /V$   (\cite{KinasiewiczPhd}, \cite{PRD2006}); in order to show that  one should  invoke  assumptions concerning boundary conditions $U^2_\infty \ll {M\over R_\infty }\ll a^2_\infty $. The calculation is long but straightforward. Thus, we finally  obtain $L_{GR}/L_N=F$; the ratio of luminosities is independent of the fraction of mass carried by the gas, in the regime of low luminosities. This means that the appropriate information on the ratio of the relativistic and Newtonian luminosities, $L_{GR}/L_N$, can be obtained just by the analysis of  accreting systems with test gas (and for these see, for instance, results in \cite{EM99} and \cite{Kinasiewicz}). This is despite the fact that  actual values of both luminosities taken separately depend on the contribution  of  the gas to total mass. We already know   that in accretion without radiation  the mass accretion rates are maximal when  $m_\ast = 2 M / 3$ and they tend to zero at both ends: i) $m_\ast \to M$ (when the density $\rho_\infty $ tends to zero) and ii) $m_\ast / M \to 0$ (when the mass of the core is negligible in comparison to the mass of the fluid) \cite{PRD2006}. That implies, for weakly radiating systems,  that luminosities behave in a similar way. But still their ratio is constant and independent of the parameter  $y$.

 \subsection{High luminosities.}

The  argument of the former subsection does not apply to  accreting luminous systems, when the total luminosity $L_0$ is  close to the Eddington limit or more generally -- when the (XY) condition is broken. That this can happen, is easily illustrated by the Newtonian model. In this model one   obtains  an equation that relates luminosity $x$ and the mass content $y$ (see Eq. (21) in \cite{AA}, in units adapted to the convention of this paper):
\begin{equation}
x = \alpha    \frac{\chi_\infty M^2}{4a^3_\infty} (1-y)
\left(y - x  \right)^2
\left( \frac{2}{5 - 3 \Gamma} \right)^{(5 - 3 \Gamma)/(2(\Gamma - 1))}.
\label{20}
\end{equation}
Here $\chi_\infty $ is a constant.  The argument that was used above relied   on the fact that the right hand side of (\ref{20}) does not depend on $x$ if $x\ll y$. But if $x$  is relatively large, then   $\dot M$ becomes $x$-dependent, and (\ref{20}) yields a relation $x=x(y)$. A similar reasoning  can also be applied  to the general relativistic model; again $\dot M$  depends on $x$  if $x$ is large. In conclusion, for luminous systems the ratio of $L_{GR}/L_N$ can become $x$ and $y$-dependent.  
 
Luminous systems are characterized by small values of the asymptotic speed of sound, $a_\infty \ll 1$.   We restrict our attention to systems that satisfy the following

{\bf X(1-Y) condition.} We will say that an accretion system satisfies the $X(1-Y)$ condition if
 $x\gg 1-y$ and $x<y/2$.

Since $y>x$, the above implies $y>2/3$. Thus X(1-Y) selects a subclass of luminous accretion systems with moderate contribution of the gas to total mass. Luminous test fluids belong to this category.  
 
Define an auxiliary quantity
\begin{equation}
\hat L \equiv -2\dot M N \frac{\rho^E}{\rho_0} +
4\pi R^2 j N k \left( 1+ \frac{U^2}{k^2}\right) .
\label{21}
\end{equation}
$\hat L$ represents local luminosity as measured by an observer stationary at $R$ \cite{Karkowski2009}. It follows from Eq. (\ref{6}) that   at the boundary of the accretion cloud  $\hat L(R_\infty )=L_{GR}$; $\hat L(R_\infty ) $ is 
the total luminosity. By differentiating Eq. (\ref{6}) and employing Eq. (\ref{10}) one can easily derive the following equation
\begin{equation}
{d\over dR} \hat L= \dot M N\kappa j   + N {\Gamma -1\over \Gamma -1-a^2} {d\over dR} \dot M.
\label{22}
\end{equation}
We have  ${d\over dR} \dot M   = -16\pi^2 R^3{\rho_0\over k} j$  \cite{Karkowski2009}. It is convenient to replace $\dot M$ by $\dot M_\infty \equiv \dot M +\kappa \int_R^{R_\infty }dr j r^3{\rho_0\over k} $; the new object is constant and  coincides with $\dot M$ at the boundary $R_\infty $.
Then,
\begin{eqnarray}
{d\over dR} \hat L&=& \dot M_\infty  N\kappa j   + \nonumber\\
&&16\pi^2 j N \left( -R^3{\rho_0\over k} {\Gamma -1\over \Gamma -1-a^2} +\kappa \int_R^{R_\infty }dr j r^3{\rho_0\over k} \right) . \nonumber\\
&&
\label{23}
\end{eqnarray}
Now, the assumptions  $a_\infty \ll 1$  and $x<y/2$ imply that in the region extending  from the sonic point to $R_\infty $ the infall velocity $U$ is small and the position of the sonic point $R_\ast $ is large ($R_\ast \gg M$). Therefore the metric function $k$ and the lapse  $N$ are close to unity.  The argument bases on the approximate validity of Newtonian relations  (\ref{16}) at  general relativistic  sonic points.    The second term on the right hand side of (\ref{23}),
\begin{equation}
\delta L\equiv   16\pi^2 j N \left( -R^3{\rho_0\over k} {\Gamma -1\over \Gamma -1-a^2} +\kappa \int_R^{R_\infty }dr j r^3{\rho_0\over k} \right)
 \label{24}
 \end{equation}
is bounded from above by the first term on the right hand of (\ref{23}), $\delta L\le   |\dot M_\infty  N\kappa j| \times  {\left( 1-y\right) \over x}$.  Indeed, the first term on the right hand side of (\ref{23}) can be written as  $j {4\pi M L_0\over \alpha L_E}$; but that is equal to $j{4\pi Mx\over \alpha }$, which in turn is larger than $j4\pi Mx$. The first term of $\delta L$ is bounded by $12\pi |j| M_g=12\pi M(1-y)$. The second term of $\delta L$ is much smaller than the first one if  sonic radii are much larger than $M$. Therefore the conclusion follows. 

    Thence in the annular region $(R_\ast , R_\infty )$ the function  $\hat L $ satisfies with good accuracy   the differential equation
\begin{equation}
{d\over dR} \hat L \approx \dot M_\infty  \kappa  {\hat L\over 4\pi R^2},
\label{25}
\end{equation}
which is solved by $\hat L =L_0 \exp \left( -\kappa {\dot M_\infty \over 4\pi R} \right) $.   Thus we obtain the same form of a solution as in the Newtonian case.   Furthermore, one can  approximate Eq. (\ref{6}) by a suitable Newtonian model in the region $(R_\ast , R_\infty )$. Indeed, it is easy to show that $\dot M\approx \dot M_\infty$. Expanding the lapse $N$ -- keeping only the first order terms in $a^2$, $M(R)/R$ and $U^2$ --  we arrive at the Newtonian equation (\ref{13}).   That means that under adopted boundary conditions the total general relativistic model is well approximated by the Newtonian model in the annulus $(R_\ast , R_\infty )$, and thus by Eq. (\ref{20}). But Eq. (\ref{20}) has a unique solution, assuming $x<y<1$. Therefore the  Newtonian limit of the general relativistic model and the Newtonian solution do coincide  and the ratio $L_{GR}/L_N$ is not only constant, but it is equal to 1.  
 
\section{Numerics.}

We compare two accreting systems, a Newtonian one and its general relativistic counterpart, that have indentical sizes, the same asymptotic masses and identical masses of compact cores, equal asymptotic temperature and  the same binding energy. Thus, it is legitimate to say that the boundary data  are ultimately: the asymptotic mass, the  mass of the core, the binding energy per unit mass $\alpha =|\phi \left( R_0\right) |$, the asymptotic speed of sound $a_\infty $ and  the  size of the system   $R_\infty  $. The total luminosity $L_0$ is not part of these data but is the sought result of the  two models.  

It appears convenient in numerical calculations to specify temporarily $\rho_{0\infty }$  and $L_0$ instead of the mass of the core. Conceptually the computational technique is the same in the two models. For a given $\rho_{0\infty}$ one randomly  chooses $L_0$ (equivalently one could choose an accretion rate, due to relation $\dot{M} = L_0 / \alpha$). This choice completely specifies $L(r)$ in the Newtonian model -- see formula (\ref{14}). Asymptotic radiation data for the general relativistic  system in turn are given by $j_\infty = \rho^E_\infty = L_0/\left( 4\pi R^2_\infty \right)$, and the mass accretion rate $\dot M_\infty  = L_0/\alpha$. During the numerical integration one    gets a subsonic solution (if the chosen $L_0$ is smaller than a critical luminosity) or  finds no solution at all (if $L_0$ is greater than a critical value). Using the bisection method one  finds this critical luminosity for which the gas flow becomes transonic. The mass of the core results from computations. Notice, that for a given $\rho_{0\infty}$ masses of the core usually differ in the Newtonian and the general relativistic models. The difference is particularly noticeable for high asymptotic sound speeds  $a_\infty $.  One should change the value of $\rho_{0\infty }$  and repeat the procedure until finally the masses of both cores are the same for both critical flows.

In this way   one  obtains a boundary of the solution set (in the plane $L_0$ -- $ M_{core}$) that consists exclusively of transonic solutions, if the mean free path of photons  is larger than the size of the system $R_\infty $.  

From a mathematical point of view we have a system of ordinary first order differential equations. The general relativistic problem also includes   the integro-algebraic constraint Eq. (\ref{6}). Numerical calculations start from the values adopted at the outer boundary $R_\infty$ and continue inward  until the equality $\alpha = 1- \frac{N\left( R\right)}{k\left( R\right) } \sqrt{1-\frac{2m\left( R\right)}{R}} $ (in the GR case)  is met at some $R$; this value of the areal radius is  denoted as $R_0$ and interpreted as  the radius of the compact core of the accreting system. For the Newtonian model the calculation continues until the gravitational potential $\phi $ becomes equal to $-\alpha $. The numerical integration  employs the 8th order Runge-Kutta method \cite{RK8}. 
The main numerical difficulty is encountered in the vicinity of the sonic point. In the general relativistic case the denominator and the numerator of Eq. (\ref{17a}) vanish for $a^2 = \frac{U^2}{k^2}$. In numerical computation, the division by very small numbers may cause errors and lead to unphysical solutions, therefore a special regularization technique had to be implemented. We omit further discussion of related technicalities, but let us  mention that because of this difficulty with the sonic point  there appear small numerical errors for $M_{core} \approx 1$ and $M_{core} \approx 0.1$ (see  Fig. 3). In the Newtonian model one has to deal with the same problem.

We choose specific numerical data, but since the accreting system possesses a simple scaling property --  as discussed in one of preceding  sections --    one can   extend  the validity of all conclusions to a large family of systems with appropriately  scaled masses $M$ and sizes $R_\infty $.

We assume standard gravitational units $G=c=1$, the polytropic index $\Gamma = 3/2$, the size  $R_\infty = 10^8 \times M$ and the mass  $M= 10^{6}M_{\odot } $, where $M_{\odot}$ is the Solar mass.  In the scaling $M=1$ one gets $\kappa = 3.6258  \times 10^{22} \left( M_\odot  /M\right)$, that is $\kappa =3.6258 \times 10^{16}$.    The Eddington luminosity
reads $L_E = 3.4658  \times 10^{-22}{M / M_\odot  }= 3.4658 \times 10^{-16}$. 
These data are  arranged  to ensure the validity of the thin gas approximation. The optical thickness (\ref{thin}) is always smaller than 1.

%%%%%%%%%%%%%%%%%%%%%%%%%%%%%%%%%%%%%%%%%%%%%%%%%%%%%%%%%%%%%%%%%%%%%%%%
\begin{figure}[h]
\includegraphics[width=\linewidth]{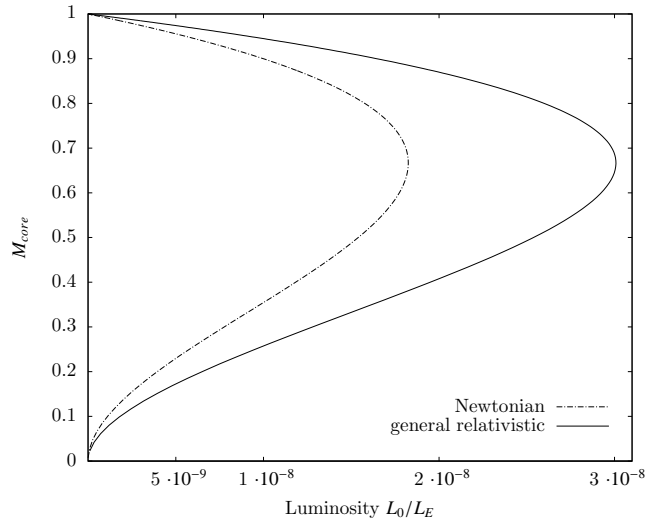}
\caption{\label{fig:1} Luminosity of general relativistic and Newtonian models. $\alpha = 0.9$ and  $a^2_\infty =10^{-1} $. The abscissa shows the
luminosity in terms of the Eddington luminosity $L_E$ and the ordinate shows the mass of the compact core.}
\end{figure}
%%%%%%%%%%%%%%%%%%%%%%%%%%%%%%%%%%%%%%%%%%%%%%%%%%%%%%%%%%%%%%%%%%%%%%%%s
\section{Results}

\begin{figure}[h]
\includegraphics[width=\linewidth]{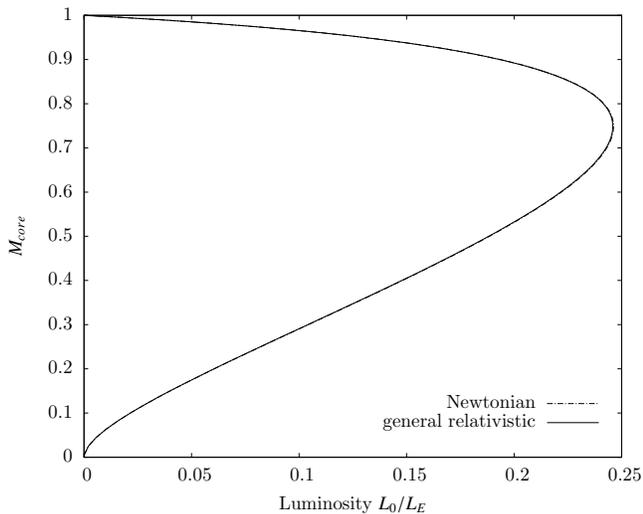}
\caption{\label{fig:2} Luminosity of general relativistic and Newtonian models. $\alpha = 0.9$ and  $a^2_\infty =10^{-6} $. The abscissa   and  ordinate are as in Fig. 1.}
\end{figure}

Figures \ref{fig:1}--\ref{fig:2} show accreting solutions on the  luminosity-(mass of the central core) diagram for 
$\alpha =0.9$.    The squared speed of sound is   $a^2_\infty = 10^{-1}$ and $a^2_\infty =10^{-6}$, respectively.  The   two figures   show transonic solution sets for the Newtonian  and general relativistic  models; they are depicted by     dashed and solid lines, respectively.    Comparing these   figures, we notice that the brightness of a  system increases sharply with the decrease of $a^2_\infty $. In the case illustrated in the second figure, maximal luminosities go up to one quarter of the Eddington luminosity.
%%%%%%%%%%%%%%%%%%%%%%%%%%%%%%%%%%%%%%%%%%%%%%%%%%%%%%%%%%%%%%%%%%%%%%%%
%%%%%%%%%%%%%%%%%%%%%%%%%%%%%%%%%%%%%%%%%%%%%%%%%%%%%%%%%%%%%%%%%%%%%%%%%
In the test gas limit, the interaction between gas and radiation is negligible and the gas accretion can be approximated by the purely hydrodynamic description. Such a case  was already analyzed in  \cite{EM99}, with the same conclusion as suggested by the comparison of  Figs 1 and 2: the larger  the asymptotic speed of sound, the larger  the  gap between the general relativistic and the Newtonian predictions. The general relativistic model  gives  significantly larger accretion rates for high asymptotic temperatures. These figures   clearly demonstrate  that luminosities depend on the fraction of mass deposited in the gas and  become maximal when this fraction is not bigger than 1/3.   Again, this aspect of the description of the regime of weakly radiating sources  agrees with the purely hydrodynamic study of \cite{PRD2006}. The position of this maximum depends weakly on the relative luminosity $L/L_E$ and it shifts   from  $y=2/3$ in Fig. 1 towards $y=0.75$ in Fig. 2.   It is clear that this effect is due to the influence of the radiation; the higher the luminosity, the larger  the mass of the core at the maximum. 

 Fig. 3   reveals  a  feature of spherical accretion that confirms the analytic proof (made in one of preceding sections), that $L_{GR}/L_N$ should be constant, at least for small luminosities. While each individual quantity $L_{GR}, L_N$ depends on the contribution  of the gas to the total mass,  their ratio is roughly constant at a given asymptotic  temperature.  In the six sets of transonic flows (lines 2-7 on Fig. 4) the ratio $L_{GR}/L_N$ is independent of the mass of accreting gas. We would like to call the reader's attention 
to  line  2, where the maximal value of $x=1/4$ is achieved at $y=0.75$ (see Fig. 2). Thus  the maximal value  of $x/y\approx 1/3$, $x\approx 1-y$ and still the fraction $L_{GR}/L_N$ is constant. This  agrees  with the analytic result shown in the second part of Section V, although the proof of this  requires $x\gg 1-y$. That suggests that analytic results can be proven under less stringent conditions than stated in Sec. V. 
  
%%%%%%%%%%%%%%%%%%%%%%%%%%%%%%%%%%%%%%%%%%%%%%%%%%%%%%%%%%%%%%%%%%%%%%%%
\begin{figure}[h]
\includegraphics[width=\linewidth ]{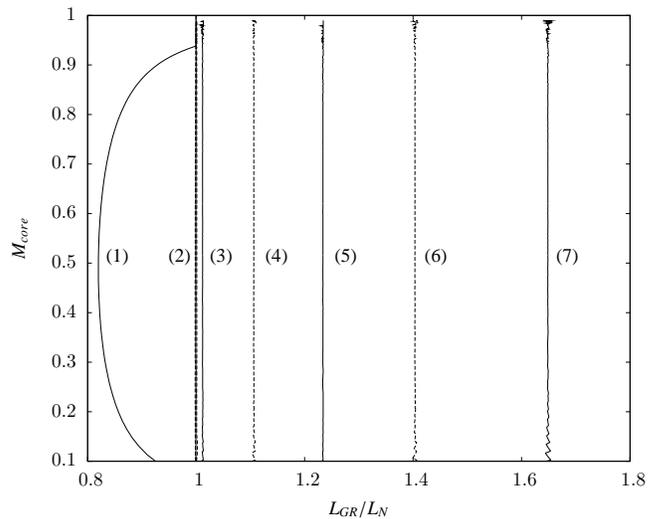}
\caption{\label{fig:3}Binding energy $\alpha = 0.9$.  The values of $L_{GR}/L_N$ are shown on the abscissa. The mass fraction $y$ is put on the ordinate. Asymptotic squared speeds of sound are: $10^{-7}$ (line no 1);  $10^{-6}, 10^{-5}, 10^{-4}$ (the three  close lines are grouped as line 2); $10^{-3}$ (line 3);
$10^{-2}$ (line 4),   $2.5 \times  10^{-2}$ (line 5), $5 \times  10^{-2}$ (line 6), $10^{-1}$ (line 7).}
\end{figure}
%%%%%%%%%%%%%%%%%%%%%%%%%%%%%%%%%%%%%%%%%%%%%%%%%%%%%%%%%%%%%%%%%%%%%%%% 
Line  1 in Fig. 3 and Fig 4. display data where the backreaction effect causes $L_{GR}/L_N$ to vary (and in particular, $L_{GR}/L_N$ can be made significantly smaller than 1). Notice, however, that in the general relativistic model  the flows cease to be transonic for $y<0.94$.  In contrast, they are always transonic in the Newtonian model. There is a small segment just below $y=1$, where the {\bf X(1-Y)} condition is met and the ratio of luminosities equals 1.  
%%%%%%%%%%%%%%%%%%%%%%%%%%%%%%%%%%%%%%%%%%%%%%%%%%%%%%%%%%%%%%%%%%%%%%%%
\begin{figure}[h]
\includegraphics[width=\linewidth ]{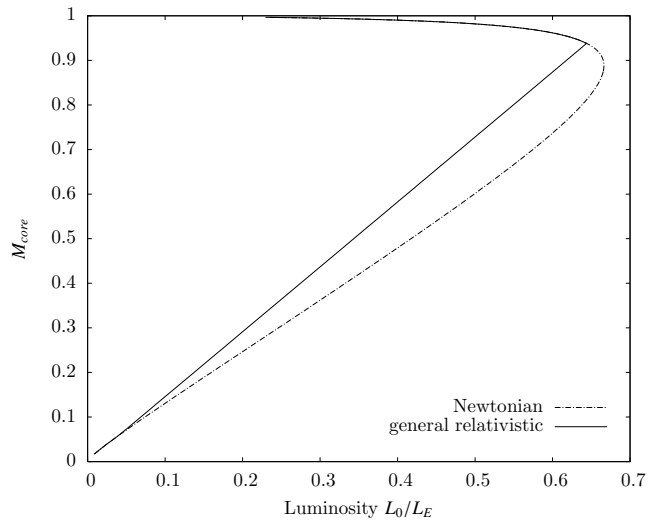}
\caption{\label{fig:4}Binding energy $\alpha = 0.9$.  The luminosities are shown on the abscissa while the mass fraction $y$ (the core mass) is put on the ordinate. Here $a_\infty^2 =10^{-7}$.  }
\end{figure}
%%%%%%%%%%%%%%%%%%%%%%%%%%%%%%%%%%%%%%%%%%%%%%%%%%%%%%%%%%%%%%%%%%%%%%%% 

\section{Conclusions}

There are interesting universal properties hidden in generalizations of the classical Bondi accretion model. It is already known that when  radiation is absent, transonic flows (that maximize mass accretion rates) correspond to the case when $y\equiv m_*/M=2/3$, irrespective of the equation  of state and the asymptotic speed of sound \cite{PRD2006}, \cite{KinasiewiczPhd}. The mass of the core is about 2/3 of the total mass of an accreting system. This paper deals with radiating accretion flows. We compare luminosities corresponding to transonic solutions of the general relativistic  and Newtonian   accretion models, assuming the same polytropic  equation of state and identical boundary data --  asymptotic speed of sound $a_\infty $, size $R_\infty $,  total (asymptotic) mass and  fraction $1-y$ of the total mass contributed by gas.  We focus our attention on the investigation of the relation between their relative  luminosity ($L_{GR}/L_N$) and   $y$. When accreting systems are characterized by low luminosity and the condition (XY) of Section V holds true (that is $L_{GR}\ll L_E \times y$ and $L_{N}\ll L_E\times y$)  then the ratio $L_{GR}/L_N$ is independent of $y$ and can be significantly larger than 1. We have found an example with  the largest value of $L_{GR}/L_N$ exceeding 1.6, but in  earlier investigation of test fluids with the polytropic index close to 5/3 the ratio of mass accretion rates $\dot M_{GR}/\dot M_N$ exceeded 10 \cite{Kinasiewicz}, which suggests that $L_{GR}/L_N$ can grow by one order of magnitude. On the other hand, when the condition X(1-Y) of Section V is valid (that is, a transonic flow is highly luminous, $x\equiv L_0/L_E\gg 1-y$ and $x<y/2$, but the contribution of gas to the mass is small), then $L_{GR}/L_N\approx 1$. These properties of  the ratio $L_{GR}/L_N$ have been derived analytically and   confirmed (under less stringent conditions) numerically. 

Acknowledgements. Tomasz Rembiasz thanks Patryk Mach for his guidance in early calculations and  Ewald M\"{u}ller for useful remarks.


\begin{thebibliography}{99}
\bibitem{EM99} E. Malec, {\it Phys. Rev. } {\bf D60}, 104043(1999).
\bibitem{Kinasiewicz} 
B.	Kinasiewicz and  P. Mach,	{\it Acta Physica Polonica} {\bf  B38}, 39(2007); B.	Kinasiewicz and  T. Lanczewski,	{\it Acta Physica Polonica} {\bf  B36}, 1951(2005).
\bibitem[{Bondi(1952)}]{Bondi} H. Bondi,   {\it MNRAS}, {\bf 112}, 195(1952).
\bibitem{Shakura} N. I. Shakura, {\it AZh} {\bf 51}, 441(1974). 
\bibitem{Thorne} Thorne, K. S. and Zytkow, A. N., {\it ApJ}, {\bf 212},
832(1977).
\bibitem{ThorneF} K. S. Thorne, R. A. Flammang \& A. N. \.Zytkow, {\it MNRAS},
{\bf 194}, 475(1981).
\bibitem{Park} M.-G. Park \& G. S. Miller {\it ApJ}, {\bf 371}, 708(1991).
\bibitem {Rezzola} L. Rezzolla \& J. C. Miller, {\it Class. and Quantum Grav.},
{\bf 11}, 1815(1994).
\bibitem{AA} J. Karkowski, E. Malec and K. Roszkowski, {\it Astronomy and
Astrophysics}, {\bf 479}, 161(2008).
\bibitem{Karkowski2009} J. Karkowski, E. Malec, K. Roszkowski and Z. \'Swierczy\'nski, 
{\it Acta Phys. Pol.} {\bf B40}, 273(2009).
\bibitem{Mihalas} D. Mihalas and B. W. Mihalas, Foundation of Radiation
Hydrodynamics, Oxford University Press New York Oxford 1984.
\bibitem{Iriondo} M. Iriondo, E. Malec and N. O' Murchadha, {\it Phys. Rev.}
{\bf D54}, 4792(1996).
\bibitem{Mach} P. Mach, {\it Acta Phys. Pol.} {\bf B38}, 3935(2007);
B. Kinasiewicz, P. Mach and E. Malec, {\it Int. J. of Geometric Meth. in Mod.
Phys.} {\bf 4}, 197(2007).
\bibitem{KinasiewiczPhd} B. Kinasiewicz, {\it Stationary accretion in general relativistic hydrodynamics},
PhD thesis, Jagiellonian University 2007.
\bibitem{PRD2006} J. Karkowski, E. Malec, B. Kinasiewicz, P. Mach and
Z. \'Swierczy\'nski, {\it Phys. Rev.} {\bf D73}, 021503(R)(2006).
\bibitem{RK8}  E. Hairer, S.P. Norsett and G. Wanner, Solving Ordinary
Differential Equations I. Nonstiff Problems. 2nd Edition, Springer Series in
Computational Mathematics, Springer-Verlag, 1993.
\bibitem{Karkowski2009a}  
J. Karkowski, E. Malec, K. Roszkowski and  Z.  \'Swierczy\'nski, {\it 	Reports on Mathematical Physics } {\bf 64}, 249(2009).
 \end{thebibliography}
\end{document}